\documentstyle[aaspp4,flushrt]{article}

\catcode`\@=11 
\def\@versim#1#2{\vcenter{\offinterlineskip
        \ialign{$\m@th#1\hfil##\hfil$\crcr#2\crcr\sim\crcr } }}
\newcommand{\Ref}{\hangindent=20pt \hangafter=1 \noindent}
\newcommand{\StartRef}{\hyphenpenalty=10000 \raggedright}

\newcommand{\beq}{\begin{equation}}
\newcommand{\eeq}{\end{equation}}
\def\lsim{\mathrel{\mathpalette\@versim<}}
\def\gsim{\mathrel{\mathpalette\@versim>}}

\def\@versim#1#2{\vcenter{\offinterlineskip
        \ialign{$\m@th#1\hfil##\hfil$\crcr#2\crcr\sim\crcr } }}
\catcode`\@=12 
\begin{document}
\title{Self-Similar Accretion Flows with Convection} 
\author{Ramesh Narayan}
\affil{Harvard-Smithsonian Center for Astrophysics, 60 Garden Street,
Cambridge MA 02138, U. S. A.}
\author{Igor V. Igumenshchev\footnote{On leave from Institute of Astronomy,
48 Pyatnitskaya St., 109017 Moscow, Russia}, 
Marek A. Abramowicz\footnote{also: Laboratorio Interdisciplanare SISSA and
ICTP, Trieste, Italy}}
\affil{Institute for Theoretical Physics, Goteborg University and Chalmers
University of Technology, 412 96 Goteborg, Sweden}
\medskip
\setcounter{footnote}{0}

\begin{abstract}
We consider height-integrated equations of an advection-dominated
accretion flow (ADAF), assuming that there is no mass outflow.  We
include convection through a mixing length formalism.  We seek
self-similar solutions in which the rotational velocity and sound speed
scale as $R^{-1/2}$, where $R$ is the radius, and consider two
limiting prescriptions for the transport of angular momentum by
convection.  In one limit, the transport occurs down the angular
velocity gradient, so convection moves angular momentum outward.  In
the other, the transport is down the specific angular momentum
gradient, so convection moves angular momentum inward.  We also
consider general prescriptions which lie in between the two limits.

When convection moves angular momentum outward, we recover the usual
self-similar solution for ADAFs in which the mass density scales as
$\rho\propto R^{-3/2}$.  When convection moves angular momentum
inward, the result depends on the viscosity coefficient $\alpha$.  If
$\alpha>\alpha_{crit1}\sim0.05$, we once again find the standard ADAF
solution.  For $\alpha<\alpha_{crit}$, however, we find a
non-accreting solution in which $\rho\propto R^{-1/2}$.  We refer to
this as a ``convective envelope'' solution or a ``convection-dominated
accretion flow.''

Two-dimensional numerical simulations of ADAFs with values of
$\alpha\lsim0.03$ have been reported by several authors.  The
simulated ADAFs exhibit convection.  By virtue of their axisymmetry,
convection in these simulations moves angular momentum inward, as we
confirm by computing the Reynolds stress.  The simulations give
$\rho\propto R^{-1/2}$, in good agreement with the convective envelope
solution.  The $R^{-1/2}$ density profile is not a consequence of mass
outflow.  The relevance of these axisymmetric low-$\alpha$ simulations
to real accretion flows is uncertain.

\noindent {\em Subject headings:} Accretion, accretion disks --- 
convection --- hydrodynamics --- turbulence
\end{abstract}

\section{Introduction}

Advection-dominated accretion flows (ADAFs, see Narayan, Mahadevan \&
Quataert 1998 and Kato, Fukue \& Mineshige 1998 for reviews) were
introduced to astrophysics by Ichimaru (1977) and have been studied
extensively in the last few years (Narayan \& Yi 1994, hereafter NY,
Abramowicz et al. 1995, Narayan \& Yi 1995a, 1995b, Chen et al. 1995).
The recent interest in ADAFs has been stimulated to a large extent by
the fact that these flows provide a natural explanation for many
phenomena associated with low-luminosity accreting black holes
(Narayan et al. 1998).  ADAFs are also relevant for understanding
black holes with super-Eddington accretion (Abramowicz et al. 1988).

NY derived an analytic self-similar solution for an ADAF which has
provided considerable insight into the properties of these flows.  A
number of authors have extended this work and obtained related
solutions (e.g. Narayan \& Yi 1995a, Honma 1996, Kato \& Nakamura
1998, Blandford \& Begelman 1999, Manmoto et al. 2000).

NY showed that the entropy of the accreting gas in an ADAF increases
towards smaller radii.  They argued that an ADAF is therefore likely
to be convectively unstable.  The convective instability was confirmed
by Igumenshchev, Chen \& Abramowicz (1996) using numerical
simulations.  More detailed simulations have been reported by
Igumenshchev \& Abramowicz (1999) and Stone, Pringle \& Begelman
(1999).

Although NY recognized the importance of convection, they incorporated
its effects only schematically in the derivation of their self-similar
solution.  We present a more detailed discussion here.  We show that,
depending on whether convective turbulence moves angular momentum
outward or inward, the nature of ADAFs may be very different.  We use
this insight to interpret the numerical simulations of Igumenshchev
and others.  A related analysis is presented in the accompanying paper
by Quataert \& Gruzinov (2000, hereafter QG).

\section{Self-Similar Scalings}

Our analysis and notation closely follow the discussion given in NY.
We work with height-integrated equations (see Narayan \& Yi 1995a for
an interpretation of height-integration).  We assume that there is no
significant mass outflow from the ADAF, i.e. that the mass accretion
rate $\dot M$ is independent of radius $R$.  The Keplerian angular
velocity is $\Omega_K=(GM/R^3)^{1/2}$ and the corresponding linear
velocity is $v_K=R\Omega_K=(GM/R)^{1/2}$, where $M$ is the mass of the
accreting star (black hole) and $R$ is the cylindrical radius.

We seek a self-similar solution which satisfies the following scalings
for the angular velocity $\Omega$, the isothermal sound speed $c_s$
and the scale height $H$:
$$
\Omega(R)=\Omega_0\Omega_K\propto R^{-3/2},
$$
$$
c_s^2(R)=c_0^2v_K^2\propto R^{-1},
$$
$$
H(R)=c_s/\Omega_K=c_0R,
$$
where $\Omega_0$ and $c_0$ are dimensionless constants to be
determined.  We write the density $\rho$ as
$$
\rho(r)=\rho_0R^{-a},
$$
so that the pressure scales as
$$
p(R)=\rho c_s^2 \propto R^{-1-a}.  
$$ 
The index $a$ is equal to 3/2 in the original self-similar solution of
NY; we reproduce this scaling in \S4.  However, as we show in \S\S5.2,
6, under appropriate conditions a very different solution is possible,
which has $a=1/2$.

As in NY, we apply the conservation laws of mass, radial momentum,
angular momentum and energy to solve for the various unknowns.  Mass
conservation requires that $\rho vRH$ be independent of $R$, where $v$
is the radial velocity.  This gives the following scaling for $v$:
$$
v(R)\propto R^{a-2}\propto R^{a-3/2}v_K.
$$

In the radial momentum equation, we assume for simplicity that $v^2\ll
v_K^2$, which corresponds to the condition that the Shakura-Sunyaev
viscosity coefficient $\alpha$ is small: $\alpha^2\ll1$.  This allows
us to ignore the ram pressure term $vdv/dR$.  We also ignore the
gradient of the turbulent pressure.  The radial momentum equation then
simplifies to a simple balance between gravity, centrifugal force and
thermal pressure gradient.  This gives
$$
\Omega_K^2R-\Omega^2R=-{1\over \rho}{dp\over dR}=(a+1)c_0^2{v_K^2\over R},
$$
which leads to the condition
$$
\Omega_0^2=1-(a+1)c_0^2. \eqno (1)
$$
Since we require $\Omega_0^2\geq0$, we see that $c_0^2\leq 2/5$ for
$a=3/2$ and $c_0^2\leq 2/3$ for $a=1/2$.

\section{Mixing Length Convection}

We follow the model of mixing length convection developed by Grossman,
Narayan \& Arnett (1993, hereafter GNA) and use their notation, except
that we replace the temperature by the isothermal sound speed $c_s^2
=kT/\mu m_p$, where $\mu$ is the dimensionless molecular weight.
The properties of convection depend sensitively on the superadiabatic
gradient
$$
\Delta\nabla c_s^2=-c_s^2{d\over dR}\ln\left(
{p^{1/\gamma}\over\rho}\right),
$$
where $\gamma$ is the adiabatic index of the gas.

To fix ideas, we consider first a non-rotating gas.  The entropy
gradient of the gas can be written in terms of $\Delta\nabla c_s^2$:
$$
T{ds\over dR}={1\over\gamma-1}{dc_s^2\over dR}-{c_s^2\over\rho}
{d\rho\over dR} = -c_p\Delta\nabla c_s^2,
$$
where $c_p=\gamma/(\gamma-1)$ is the specific heat at constant
pressure (in units of $k/\mu m_p$).  The Brunt-V\"ais\"al\"a frequency
$N$ is given by
$$
N^2=-{1\over\rho}{dp\over dR}{d\over dR}\ln\left({p^{1/\gamma}\over\rho}
\right)
=-{g_{eff}\over c_s^2}\Delta\nabla c_s^2,
$$
where $g_{eff}\equiv -(1/\rho)(dp/dR)$ is the radial effective
gravity.  When $N^2$ is positive, perturbations in the gas have an
oscillatory behavior and the medium is convectively stable.  However,
when $N^2$ is negative, i.e. when $N$ is imaginary, perturbations have
a runaway growth, leading to convection.  Convection is present
whenever $\Delta\nabla c_s^2$ is positive, i.e. when the entropy
decreases outward.  This is the well-known Schwarzschild criterion.

We assume that all mixing lengths are equal to a single length $L_M$,
and we set the dimensionless thermal expansion coefficient to unity
(ideal gas).  We also ignore the effect of the microscopic viscosity
$\nu$ on the motion of convective eddies.  This is not a very safe
assumption, but it simplifies the analysis considerably.  We believe
that it will not affect the qualitative nature of the results.  With
these asumptions, the various coefficients, $A$, $B$, $D$, $E$, in GNA
simplify to $A=D=0$, $B=E=(2/L_M)$.

Borrowing results from GNA, the root mean square turbulent velocity of
convective blobs is given by
$$
\sigma=\left({g_{eff}\Delta\nabla c_s^2\over BEc_s^2}\right)^{1/2}
={L_M\over 2}(-N^2)^{1/2},
$$
and the convective energy flux is
$$
F_c={\rho c_pc_s^2\over g_{eff}}(A+B\sigma)\sigma=
-{L_M^2(-N^2)^{1/2}\over 4}\rho T{ds\over dR}.
$$
Let us define the effective diffusion constant $K_c$ for convective
energy transport by the relation
$$
F_c=-K_c\rho T{ds\over dR}.
$$
We then obtain
$$
K_c={L_M^2\over 4}(-N^2)^{1/2}. \eqno (2)
$$
We see that $K_c$ has an intuitively obvious form; it is the product
of $(-N^2)^{1/2}$, which describes the characteristic frequency
associated with convective motions (more precisely, it is the growth
rate of perturbations), and the square of $L_M$, the characteristic
length scale of convection.  The standard treatment of mixing length
convection in astrophysics gives a coefficient of $1/4\sqrt{2}$ rather
than 1/4 (e.g. Cox \& Giuli 1968, Kippenhahn \& Weigert 1990).  Since
empirical estimates of the mixing length make use of the latter
coefficient, we incorporate this difference in our definition of the
dimensionless mixing length $l_M$ below.

All mixing lengths are equal in our treatment.  It is therefore
natural to assume that all transport phenomena in the convective
medium have diffusion constants of the same order as equation (2).  In
this spirit, we assume that the above expression for $K_c$ is valid
also for convective angular momentum transport.  Of course, there
could be differences in the relative efficiencies of energy and
angular momentum transport, associated for instance with different
mixing lengths for the two phenomena.  As we highlight in this paper,
there are large uncertainties even with regard to the sign of the
angular momentum flux.

\subsection{Convection in a Differentially-Rotating ADAF}

When there is rotation, new frequencies enter the problem and
convection is no longer determined purely by the Brunt-V\"ais\"al\"a
frequency $N$.  In the spirit of our height-integrated approach, let
us consider displacements of gas elements in the equatorial plane.  In
this case, the effective frequency $N_{eff}$ of convective blobs is
given by
$$
N_{eff}^2=N^2+\kappa^2, \eqno (3)
$$
where $\kappa$ is the epicyclic frequency for particle motions in the
equatorial plane; for $\Omega\propto R^{-3/2}$, we have
$\kappa=\Omega$.  By analogy with equation (2), we may then
write the diffusion constant $K_c$ as
$$
K_c={L_M^2\over 4}(-N_{eff}^2)^{1/2}. \eqno (4)
$$

QG have shown that the restriction to equatorial displacements leads
to an underestimate of the growthrate of convective modes.  In the
Appendix, we present a more general analysis in which we remove the
restriction to equatorial displacements.  In the rest of the paper,
however, we work with equations (3) and (4).

For a self-similar ADAF, we find that
$$
N^2=-{1\over\rho}{dp\over dR}{d\over
dR}\ln\left({p^{1/\gamma}\over\rho} \right)=
-{(1+a)[1-(\gamma-1)a]\over\gamma}c_0^2\Omega_K^2,
$$
$$
N_{eff}^2=\left\{-{(1+a)[(\gamma+1)-(\gamma-1)a]\over\gamma}c_0^2
+1\right\}\Omega_K^2.  
$$ 
It is not immediately obvious from the above expression whether a
given flow will be convectively stable or unstable, since that depends
on the sign of $N_{eff}^2$ which cannot be determined until the full
problem is solved and the value of $c_0^2$ is obtained.

If $N_{eff}^2<0$, we have a convectively unstable medium and the
diffusion constant for energy transport is given by equation (4).  Let
us write the mixing length $L_M$ in terms of the pressure scale
height:
$$
L_M=2^{-1/4}l_MH_p, \qquad H_p=-{dR\over d\ln p}={R\over(1+a)},
$$
where $l_M$ is the usual dimensionless mixing length parameter (called
$\alpha$ in the solar and stellar convection literature, cf. Kim et
al. 1996, Abbett et al. 1997).  The additional factor of $2^{-1/4}$
has been introduced in order to bring our formulae in line with the
usual versions of mixing length convection (cf the discussion below eq
2).  We also write the diffusion constant in the usual form
$$
K_c=\alpha_c{c_s^2\over\Omega_K},
$$
where $\alpha_c$ is a dimensionless coefficient which describes the
strength of convective diffusion; this coefficient is similar to the
usual Shakura \& Sunyaev $\alpha$ which is used to parameterise
the strength of viscosity.  We then find that
$$
\alpha_c={l_M^2\over4\sqrt{2}(1+a)^2c_0^2}\left\{{(1+a)[(\gamma+1)-
(\gamma-1)a]\over\gamma}c_0^2-1\right\}^{1/2}, \eqno (5)
$$
and the convective energy flux takes the form
$$
F_c=-\alpha_c{c_s^2\over\Omega_K}\rho T{ds\over dR}.
$$

In the Sun, the mixing length $l_M$ is estimated to be approximately
$\sim1.5$ (e.g. Abbett et al. 1997).  However, it is believed that
$l_M$ should be smaller when there is a strong superadiabatic gradient
(e.g. Kim et al. 1996); for instance, Asida (1999) estimates
$l_M\sim1.4$ in red giant envelopes.  In view of the relatively large
superadiabatic gradient present in ADAFs, the convection in these
flows is likely to be more similar to that found in red giant
envelopes than in the Sun's convection zone.  So we may expect
$l_M^2\sim2$ in ADAFs.  We use this value for numerical estimates, but
we note that there is some uncertainty in the estimate.

Transport of angular momentum by convection is a complex subject and
there is no consensus on how it operates.  We consider two extreme
possibilities.  In one limit we assume, following NY, that convection
behaves like normal viscosity; that is, we write the flux of angular
momentum as
$$
\dot J_c=-\alpha_c{c_s^2\over\Omega_K} \rho R^3{d\Omega\over dR}.
\eqno (6)
$$
This corresponds to the assumption that the convective angular
momentum flux is oriented down the {\it angular velocity gradient},
i.e. that convection tries to drive a system towards a state of
uniform rotation, just as microscopic viscosity does.  For
$\Omega\propto R^{-3/2}$, it corresponds to outward transport of
angular momentum.  Equation (6) also assumes that the diffusion
constant for angular momentum transport is equal to that for energy
transport, hence the use of the same constant $\alpha_c$.

An alternative possibility is that the convective flux scales as
$$
\dot J_c=-\alpha_c{c_s^2\over\Omega_K} \rho R{d(\Omega R^2)\over dR}.
\eqno (7)
$$
This means that the convective angular momentum flux is oriented down
the {\it specific angular momentum gradient}, i.e. that convection
tries to drive a system towards a state of uniform specific angular
momentum.  For $\Omega\propto R^{-3/2}$, it corresponds to inward
transport of angular momentum.

We concentrate on these two limiting cases in \S4 and \S5.  However,
as Kumar, Narayan \& Loeb (1995) showed, there could be a continuum of
intermediate possibilities, depending on the nature of the
interactions between convective eddies.  Therefore, we discuss in \S6
the general case where the angular momentum flux takes the form
$$
\dot J_c=-\alpha_c{c_s^2\over\Omega_K} \rho R^{3(1+g)/2}
{d[\Omega R^{3(1-g)/2}]\over dR}.
\eqno (8)
$$
Here, the index $g$ allows us to tune the physics of convective
angular momentum transport.  When $g=1$, we reproduce equation (6) and
when $g=-1/3$ we reproduce equation (7).  In principle, any value of
$g$ between these two extremes is possible.  In our problem, the
specific case $g=0$ corresponds to zero angular momentum transport.

\section{Outward Angular Momentum Transfer}

We assume that convection moves angular momentum outward according to
equation (6).  In this case, the only valid self-similar solution is
the one found by Narayan \& Yi (1994), which has $a=3/2$, $\rho\propto
R^{-3/2}$.  For this case, we may write the radial velocity as
$$
v(R)=v_0v_K\propto R^{-1/2},
$$
where $v_0$ is a constant.

We consider first the angular momentum equation.  We follow Narayan \&
Yi (1994) and look for a self-similar solution in which the net flux
of angular momentum vanishes.  This implies that the sum of the
angular momentum fluxes due to viscosity $\dot J_v$, convection $\dot
J_c$ and advection $\dot J_{adv}$ is equal to zero.  Thus
$$
\dot J=\dot J_v+\dot J_c+\dot J_{adv}=
-(\alpha+\alpha_c){c_s^2\over \Omega_K}\rho R^3{d\Omega\over dR}
+\rho Rv\Omega R^2=0,
$$
which simplifies to
$$
v_0=-{3\over 2}(\alpha+\alpha_c)c_0^2. \eqno (9)
$$

We consider next the energy equation.  Ignoring cooling, but including
energy transport by convection, this equation takes the form
$$
\rho vT{ds\over dR}+{1\over R^2}{d\over dR}(R^2F_c)=Q^+.
$$
The viscous dissipation rate is equal to (shear stress)$\times$(rate
of strain):
$$
Q^+ =(\alpha+\alpha_c){c_s^2\over \Omega_K}\rho R^2
\left({d\Omega\over dR} \right)^2.
$$
Substituting equation (1) for $\Omega_0$ and equation (8) for $v_0$,
and simplifying, this reduces to the following relation,
$$
\left[{(5-3\gamma)\over(\gamma-1)}(3\alpha+\alpha_c)+{45\over 2}
(\alpha+\alpha_c)\right]c_0^2=9(\alpha+\alpha_c), \eqno (10)
$$
where we recall that $\alpha_c$ is itself a function of $c_0^2$
(cf. eq. 5).  We note that equation (10) is equivalent to equation
(36) in Kato \& Nakamura (1998).  Given the values of $\alpha$ and
$\gamma$, this equation may be solved for $c_0^2$.  Equations (1) and
(9) then allow us to calculate the angular velocity and radial
velocity, thus completing the solution.

The solid lines in Fig. 1 show the variation of the convective
coefficient $\alpha_c$ as a function of the viscous coefficient
$\alpha$ for three choices of $\gamma$: 1.6, 1.5, 1.4.  We see that
$\alpha_c$ lies in the range 0.02 to 0.07, depending on the values of
$\alpha$ and $\gamma$.  The estimate of $\alpha_c$ is directly
proportional to $l_M^2$, which we have set equal to 2.

An important result of this analysis is that we obtain a consistent
solution for all values of $\alpha$ and for any $\gamma<5/3$.  In
other words, the NY self-similar solution, with $a=3/2$, is valid in
the presence of convection for all values of $\alpha$, provided
convection transfers angular momentum outward with the same efficiency
as it transfers energy (i.e. same $\alpha_c$).  This result was
already obtained by NY using a somewhat simplified model of
convection.  We have presented here a more detailed analysis, using a
slightly more rigorous version of mixing length convection.

Note that for the NY self-similar solution the value $\gamma=5/3$ is
special.  For $\gamma=5/3$ and any value of $\alpha$, we find
$c_0^2=2/5$, $\Omega_0^2=0$ (no rotation) and $\alpha_c=0$ (no
convection).  There are no self-similar solutions of the NY form for
$\gamma>5/3$.

\section{Inward Angular Momentum Transfer}

We now consider the case when angular momentum is moved inward by
convection, following the prescription given in equation (7).  To our
knowledge, this case has not been considered previously in the theory
of ADAFs.

\subsection{Self-Similar ADAF Solution}

We consider first the case when $a=3/2$.  The analysis proceeds very
similarly as in the previous section.  The angular momentum equation
looks the same as before, except that $\dot J_c$ is now proportional
to $-d(\Omega R^2)/dR$ rather than $-d\Omega/dR$.  Correspondingly,
the vanishing of the angular momentum flux gives the following
condition:
$$
v_0=-{1\over2}(3\alpha-\alpha_c)c_0^2. \eqno (11)
$$

In the energy equation, we note that the net viscous stress, including
the effect of viscosity, is now proportional to
$(3\alpha-\alpha_c)/2$.  We once again write $Q^+$ as (shear
stress)$\times$(rate of strain).  Then, instead of equation (10), we
obtain the following relation,
$$
{(5-3\gamma)\over(\gamma-1)}(\alpha-\alpha_c)c_0^2=(3\alpha-\alpha_c)
\left(1-{5\over2}c_0^2\right), \eqno (12)
$$
where once again $\alpha_c$ is a function of $c_0^2$.

The solid lines on the right in Figure 2 show the variation of
$\alpha_c$ with $\alpha$ for various values of $\gamma$.  As before,
we see that $\alpha_c$ lies in the range $\sim0.02-0.07$.  However, we
now find that a consistent solution is available only for $\alpha$
greater than a certain critical $\alpha_{crit1}$, whose value lies in
the range 0.03 to 0.08.  The value of $\alpha_{crit1}$ depends on the
value of $\gamma$, and is also directly proportional to $l_M^2$.

When $\alpha=\alpha_{crit1}$, the sound speed takes on its maximum
value, $c_0^2=2/5$ (and the rotation goes to zero, cf. eq 1).  Knowing
this fact, it is straightforward to show that
$$
\alpha_{crit1}=\alpha_{c,crit1}={l_M^2\over 20}\left({5-3\gamma
\over\gamma}\right)^{1/2},
$$
where $\alpha_{c,crit1}$ is the value of the convective $\alpha_c$
when $\alpha=\alpha_{crit1}$.  Note once again that the solution is
valid only for $\gamma\leq5/3$.

\subsection{Self-Similar Convective Envelope Solution}

What happens if $\alpha<\alpha_{crit1}$?  In this case, there is no
NY-like self-similar solution with $a=3/2$.  A possible reason for the
lack of solution is the following.  Since advection moves angular
momentum inward and since we have assumed (in this section) that
convection also moves angular momentum inward, the only way to have a
consistent self-similar accretion solution is for $\alpha$ viscosity
to move an equivalent amount of angular momentum outward.  If the
parameter $\alpha$ is very small, the viscous flux is unable to cope
with the inward flux due to convection, and there is no consistent
accretion solution.

However, we find that when $\alpha$ is small, a completely different
solution, with $a=1/2$, $\rho\propto R^{-1/2}$, is possible.  This
is a {\it non-accreting} solution with $v=0$ (at least in the limit of
perfect self-similarity).  We refer to it as a ``convective envelope''
solution, or a ``convection-dominated accretion flow'' (QG).

Since $v=0$ in this solution, the advected angular momentum flux $\dot
J_{adv}$ vanishes.  Therefore, we must have $\dot J_v+\dot J_c=0$,
i.e. there must be a perfect balance between outward angular momentum
transport via viscosity and inward transport via convection.  This
leads to the following condition on $c_0^2$:
$$
\alpha={1\over3}\alpha_c={l_M^2\over27\sqrt{2}c_0^2}\left\{
{3(\gamma+3)\over4\gamma}c_0^2-1
\right\}^{1/2}. \eqno (13)
$$
If $\alpha$ is very small, then $\alpha_c$ is also very small.  In
this limit, we have a flow which is marginally stable to convection,
with $c_0^2=4\gamma/ 3(\gamma+3)$.  (This result is valid only if we
restrict the analysis to equatorial motions; see the Appendix for a more
general discussion.)

The convective envelope solution satisfies the energy equation
trivially.  Since the net angular momentum flux vanishes, there is no
shear stress, and $Q^+=0$.  Furthermore, since $v=0$, there is no
advection of entropy.  Finally, since $a=1/2$, the convective energy
flux $F_c\propto R^{-2}$ and therefore the divergence of convective
energy flux vanishes.  This last property, which is essential in order
to satisfy the energy equation, is what fixes $a=1/2$ for the
solution.

Equation (13) allows us to solve for $\alpha_c$ as a function of
$\alpha$ and $\gamma$.  The solid line segments on the left in Fig. 2
show the results.  Interestingly, we find that there is a consistent
solution only if $\alpha$ is less than a critical $\alpha_{crit2}$.
The critical value is obtained by setting $c_0^2$ equal to the
largest value it is allowed for $a=1/2$, viz. 2/3 (see eq 1).  This
gives
$$
\alpha_{crit2}={1\over3}\alpha_{c,crit2}={l_M^2\over 36}\left(
{3-\gamma\over\gamma}\right)^{1/2}.
$$
We see that $\alpha_{crit2}$ is of the same order as $\alpha_{crit1}$,
i.e. $\sim0.05$, but that the two are not exactly equal.  An
interesting feature is that $\gamma=5/3$ is not special for this
solution.  Consistent flows can be obtained even for larger $\gamma$.

Note that a convection-dominated accretion flow is technically not an
accretion flow at all, but a static configuration.  In practice of
course, a small amount of mass will flow into the central black hole
from the innermost region and this will drive a small amount of
accretion.  However, in contrast to the ADAF solution described by NY,
where the mass accretion rate is determined by the density and
temperature on the outside and is not sensitive to the radius of the
inner boundary, in the convective envelope the accretion rate is
determined entirely by the conditions at the inner boundary and
therefore on the location of the inner boundary.

Another interesting feature is that the convective envelope solution
has a significant outward flux of energy carried by convection.  This
energy flux is constant with radius and clearly originates near the
center.  Indeed, the flux originates from the (small amount of) mass
which is accreted at the inner edge.  A fraction of the binding energy
of this accreted mass is diverted outward by convection and is
transported to large radii, where it is presumably radiated in some
fashion.  This feature of the convective envelope solution, namely
that energy which is generated deep in the center is transported a
large distance radially before being radiated, is very similar to what
happens in the convection zone of low-mass main-sequence stars like
the Sun and red giants.

\section{General Angular Momentum Transfer}

For completeness, we briefly discuss the general case in which the
angular momentum transfer is described by equation (8) with an
arbitrary index $g$.

If we consider an NY-like self-similar solution with $a=3/2$,
equations (10) and (12) are replaced by the more general relation
$$
\left[{3(5-3\gamma)\over(\gamma-1)}c_0^2+{45\over2}c_0^2-9\right]
\alpha = \left[9g-{(3g-2)(5-3\gamma)\over(\gamma-1)}c_0^2-
{45g\over2}c_0^2\right]\alpha_c,
$$
where $\alpha_c$ is still given by equation (5).  It is easily
verified that this equation reduces to equation (10) for $g=1$ and
equation (12) for $g=-1/3$.  We can also show that the critical
$\alpha_{crit1}$ becomes
$$
\alpha_{crit1}=\left({2\over3}-g\right)\alpha_{c,crit1}=
\left({2\over3}-g\right){l_M^2\over 20}\left({5-3\gamma
\over\gamma}\right)^{1/2}.
$$
We see that $\alpha_{crit1}$ is positive whenever $g<2/3$.  This is an
interesting result.  It shows that the breakdown of the NY solution at
small $\alpha$ is not unique to the prescription (7) for the angular
momentum transfer, which corresponds to $g=-1/3$.  The same thing
happens even when there is no angular momentum transfer by convection
($g=0$) or when there is angular momentum transfer outward but with a
reduced efficiency compared to equation (6) ($0<g<2/3$).  In these
cases, the critical $\alpha_{crit1}$ is smaller than when $g=-1/3$,
which means that $\alpha$ has to be lower before the NY solution would
fail.

Similarly, when we repeat the analysis of \S5.2 that led to the
convective envelope solution, we find for a general value of $g$ that
equation (13) is replaced by
$$
\alpha=-g\alpha_c=-{gl_M^2\over9\sqrt{2}c_0^2}\left\{
{3(\gamma+3)\over4\gamma}c_0^2-1\right\}^{1/2}.
$$
We see that the convective envelope solution is allowed for any $g<0$,
but not for $g>0$.  The critical $\alpha_{crit2}$ up to which the
solution is available is
$$
\alpha_{crit2}=-g\alpha_{c,crit2}=-{gl_M^2\over 12}\left(
{3-\gamma\over\gamma}\right)^{1/2}.
$$

One interesting fact that emerges from this analysis (as well as the
simpler version presented in \S5) is that there are certain parameter
regimes for which neither an NY-like self-similar solution nor a
convective envelope solution is possible.  Any flow with $0<g<2/3$ and
$\alpha<\alpha_{crit1}$ belongs to this category.  For this parameter
range, power-law solutions, if any, would violate one of our basic
assumptions.  The most likely modification is the scaling
$\Omega\propto\Omega_K\propto R^{-3/2}$ which we have assumed.  Honma
(1996), for instance, discovered an ADAF solution for $\gamma=5/3$
which has $\Omega\propto R^{-1/2}$.  This solution has been developed
further by Kato \& Nakamura (1998) and Manmoto et al. (2000).

\section{Comparison with Numerical Simulations}

In this section we compare the theory developed in the previous
section with numerical results from a low-viscosity two-dimensional
(2D) simulation of a non-radiative accretion flow.  The model was
calculated by solving the non-relativistic time-dependent
Navier-Stokes equations for an accretion flow in the gravitational
field of a point mass.  All components of the viscous stress were
included.  The details of the numerical technique are described in
Igumenshchev \& Abramowicz (1999).

The model considered here has $\alpha=0.01$, $\gamma=5/3$, inner
boundary of the accretion flow at $R_{in}=3 R_g$ and outer boundary at
$R_{out}=8\times 10^3 R_g$, where $R_g$ is the gravitational radius of
the black hole.  Mass is steadily injected within an equatorial torus
near the outer boundary of the grid.  There is no cooling in the
accreting gas.

Initially, there is little or no mass in the grid.  As the injected
mass spreads and accretes, the mass within the grid increases.  After
a period of time comparable to the viscous time scale at the outer
radius, the accretion flow achieves a quasi-stationary behaviour, and
may be considered to be in steady state.  The ``steady state'' is,
however, only in a time-averaged sense since the flow has convective
motions which introduce chaotic fluctuations at any given point.  We
compute time-averaged properties of the flow for comparison with the
theoretical predictions by averaging over a time equal to 44 Keplerian
periods measured at $100 R_g$.

Before we describe the results we would like to emphasize an important
point.  Stone \& Balbus (1996) showed that if azimuthal pressure
gradients are absent, then turbulence in a Keplerian disk can only
transport angular momentum inward.  By symmetry, 2D axisymmetric
simulations do not have azimuthal pressure gradients.  Because of
this, each eddy preserves its angular momentum as it moves (in the
absence of ordinary viscosity).  Therefore, turbulent mixing of eddies
in axisymmetric simulations tends to drive the system towards a state
of constant specific angular momentum, and the transfer of angular
momentum behaves like equation (7).  We emphasize that this is merely
a consequence of axisymmetry, and does not necessarily represent the
properties of real convection.

Our numerical simulations confirm the above expectation for the
direction of transfer of angular momentum.  Figure~3 shows the
$(R,\phi)$-component of the Reynolds stress
$\sigma_{R\phi}\equiv\langle\rho v_R' v_\phi'\rangle$ as a function of
radius, where the stress has been averaged over polar angle $\theta$
as follows:
$$
\sigma_{R\phi}(R)={\int_0^\pi \sigma_{R\phi}(R,\theta)
      \sin\theta d\theta \over \int_0^\pi \rho(R,\theta) \sin\theta
      d\theta}.
$$
We see that, except near the boundaries ($R<5R_g$ and $R>10^3R_g$),
the Reynolds stress is negative everywhere in the flow.  This
indicates that the bulk convective motions in the gas move angular
momentum inward.  Equating the numerical estimate of $\sigma_{R\phi}$
and the convective stress $K_c\rho(\partial\Omega R^2/\partial R)/R$,
we obtain an estimate $\alpha_c\simeq 3\langle v_R'
v_\phi'\rangle/v_K^2= 10^{-2}$, where we have used $H/R=1/2$.
Thus, $\alpha_c/\alpha\simeq 1$ in the numerical
model.  

Because the convective angular momentum transfer in the simulation
follows equation (7), we must make use of the analysis presented in
\S5 for interpreting the numerical results.  Moreover, since $\alpha$
is very small ($\alpha=0.01$), we expect the simulation to reproduce
the low-$\alpha$ convective envelope solution with $\rho\propto
R^{-1/2}$ rather than the high-$\alpha$ NY branch of solution.  To
test this prediction, we show in Fig.~4 the profiles of $\rho$,
$c_s^2/v_K^2$, $|v|/v_K$ and $\Omega/\Omega_K$ as functions of $R$.
Each quantity has been averaged over $\theta$ exactly as in the case
of the Reynolds stress.

The simulation gives a density profile with $\rho \propto R^{-1/2}$,
as expected from the theory.  A similar $R^{-1/2}$ variation was seen
also in low-viscosity ADAF simulations by Igumenshchev \& Abramowicz
(1999) and by Stone et al. (1999).  The latter authors considered a
variety of viscosity prescriptions, of which Run K corresponded to the
usual $\alpha$ prescription; this run exhibited $\rho\propto R^{-1/2}$
behavior, as expected for the convective envelope.  It may be worth
emphasizing that the $R^{-1/2}$ density profile in the convective
envelope solution (which is very different from the $R^{-3/2}$ profile
of the self-similar NY solution) is not a consequence of mass loss.
In the analytic work presented in this paper, for instance, we
explicitly assume that there is no mass loss.  Both Igumenshchev \&
Abramowicz (1999) and Stone et al. (1999) found that for low values of
$\alpha$ there were no powerful unbound outflows in their numerical
simulations.

In this context, note that because the medium is convectively
turbulent and is not accreting, roughly half the mass at any radius at
any given time will be flowing in and half will be flowing out.  We
could estimate a ``mass inflow rate'' $\dot M_{in}(R)$ by adding up
all the inflowing gas elements.  This would give $\dot
M_{in}\sim4\pi\rho \sigma RH/2$ where $\sigma$ is the root mean square
velocity of the turbulent eddies.  In \S3 we showed that $\sigma$
scales as $L_M(-N_{eff}^2)^{1/2}\sim v_K \sim R^{-1/2}$.  Therefore,
for $\rho\propto R^{-1/2}$ and $H\propto R$, we find that $\dot
M_{in}\propto R$.  It might be tempting to identify the rapid decline
of $\dot M_{in}$ with decreasing radius as evidence for a massive
outflow.  But there is no reason to think that the outflowing gas at
any radius in the interior of the flow will escape and flow out to
infinity.  It is more than likely that the outflowing gas forms part
of convective eddies, so that the mass that is flowing out at a
certain time will later turn around and flow in.  If such is the case,
then the quantity $\dot M_{in}$ is not very useful.

The analytic convective envelope solution predicts that $\dot M=0$ and
$v=0$.  As Fig. 4 shows, the mean radial velocity in the simulation is
extremely small, but it is not exactly zero.  The small non-zero value
is because the simulated flow has an inner boundary where some mass
can flow into the black hole.  Thus there is a small $\dot M$ which
leads to a finite $v$.  The scalings discussed in \S2 show that for
$\rho\propto R^{-1/2}$, we expect the velocity to scale as $v\propto
v_K/R$.  This is confirmed in Fig. 4.

The profiles of $c_s^2(R)$ and $\Omega(R)$ in Fig. 4 show the
predicted power-law behaviors of these quantities for $10 R_g<R<10^3
R_g$.  However, a comparison of actual numerical values does not give
good agreement.  From equations (1) and (13), for $a=1/2$,
$\gamma=5/3$ and very small $\alpha$, the theory predicts
$$
{c_s^2\over v_K^2}={4\gamma\over 3(3+\gamma)}=0.48,\qquad\qquad
{\Omega\over\Omega_K}=\left({3-\gamma\over 3+\gamma}\right)^{1/2}=0.53.
$$
The numerical simulations give $c_s^2/v_K^2=0.26$ and
$\Omega/\Omega_K=0.9$.  As QG have shown, this discrepancy is the
result of the height-integrated approximation.  We discuss in the
Appendix an improved theory that avoids this approximation.  The
theory predicts $c_s^2/v_K^2=0.29$ and $\Omega/\Omega_K=0.75$, in
better agreement with the simulations.

An important feature of the convective envelope solution is that there
is a strong outward flux of energy $F_c$ transported by convection.
The corresponding luminosity is equal to $\dot{E}_{c}=4\pi RH F_c$,
which is independent of radius.  Comparison of the theoretical
estimate with numerical results from the simulation show quite good
agreement.  Substituting the numerical value of $\rho R^{1/2}$ at
$R=100R_g$ in the analytical expression for $F_c$, we estimate
$\dot{E}_{c}\simeq 0.2\alpha_c$ in units of $\dot{M}c^2$, where
$\dot{M}$ is the mass accretion rate into the black hole.  On the
other hand, the numerical simulations give
$\dot{E}_{c}=\dot{E}_{tot}-\dot{E}_{v}\simeq 2\times 10^{-3}$, where
$\dot{E}_{tot}$ is the total outward energy flux and $\dot{E}_{v}$ is
the energy flux due to viscosity.  Equating the two estimates of the
convective flux we get $\alpha_c\simeq 10^{-2}$.  This estimate of
$\alpha_c$ agrees with our previous estimate from the Reynolds stress,
which is encouraging considering the uncertainties (physical and
numerical) of the two estimates.  Thus, we feel that the convective
envelope model is a good description of numerically simulated 2D ADAFs
at low $\alpha$.

\section{Discussion}

The main result of this paper is that, in addition to the NY
self-similar ADAF solution which has density $\rho\propto R^{-3/2}$,
there is a second power-law solution, a radiatively inefficient
``convective envelope'' solution or ``convection-dominated accretion
flow,'' in which $\rho\propto R^{-1/2}$.  The second solution
corresponds to a static envelope in which the mass accretion rate is
very small.  Indeed, $\dot M$ vanishes in the limit when the inner
edge of the accretion flow moves down to 0.  The solution has
well-developed convection and is perfectly balanced in the sense that
it transfers as much angular momentum outward via viscosity as it
transports inward via convection.  There is, however, a net outward
flux of energy.

The convective envelope solution is possible only when two
conditions are simultaneously met:

\noindent (i) The viscosity coefficient $\alpha$ must be fairly small.
For a reasonable choice of the mixing length parameter ($l_M^2\sim2$),
we find that we require $\alpha<\alpha_{crit2}\sim 0.05$ (see Fig. 2).
The improved analysis presented in the Appendix modifies the value
upward.

\noindent (ii) Convection must transport angular momentum inward
($g=-1/3$, \S5) or not transport angular momentum at all ($g=0$, \S6).
If it transports angular momentum outward, it must do so with less
efficiency than it transports energy; specifically, we require $g<2/3$
(\S6).

The numerical simulations discussed in \S7 and in Igumenshchev \&
Abramowicz (1999) and Stone et al. (1999) satisfy both of the above
conditions.  In particular, condition (ii) is automatically satisfied
by all numerical simulations of ADAFs carried out so far since all
these simulations have been axisymmetric.  The axisymmetry guarantees
that there are no azimuthal pressure gradients and so convective
eddies transport angular momentum inward (Stone \& Balbus 1996).  We
find that the numerical results agree quite well with the analytical
results (Figs. 3, 4, \S7).

In all cases where either condition 1 or 2 is violated, the only
self-similar solution we have been able to find is the solution
derived by NY, in which $\rho\propto R^{-3/2}$.  For given outer
boundary conditions, this solution leads to a considerably larger mass
accretion rate than that obtained with the convective envelope
solution.

Igumenshchev \& Abramowicz (1999) found that convection plays a
negligible role when $\alpha$ is $\gsim 0.1$.  This is consistent with
our analysis, which shows that the convective $\alpha_c$ is much less
than $\alpha$ in this regime (Figs. 1, 2).  Indeed, our analysis
overestimates $\alpha_c$ for such models because the mixing length
formalism we employ is a local steady state theory which assumes that
convection has achieved saturation amplitude.  However, the convective
turnover time, which is given by $t_{conv}=(-N_{eff}^2)^{-1/2}$ is of
order $1/\alpha_c\Omega_K$, whereas the accretion time in the NY
solution is of order $t_{acc}\sim1/\alpha\Omega_K$.  Thus
$t_{conv}/t_{acc}\sim\alpha/\alpha_c$, and this becomes large as
$\alpha$ increases.  The net effect would be to reduce $\alpha_c$
below the values shown in Figs. 1 and 2.  Thus, we expect convective
motions to be negligibly small for $\alpha\gsim0.1$, as confirmed by
Igumenshchev \& Abramowicz (1999).  (Stone et al. 1999 did not explore
models with such large values of $\alpha$.)

The range $\alpha\sim0.1-0.3$ is found to be very interesting in
Igumenshchev \& Abramowicz's (1999) simulations.  For $\alpha=0.3$,
$\gamma=1.5$, they found that there is a pure inflow, while for other
choices, e.g. $\alpha=1$, $\gamma=1.5$, they obtained a stable bipolar
outflow.  Igumenshchev (2000) studied a model with $\alpha=0.1$,
$\gamma=5/3$ and found a global meridional circulation pattern
accompanied by a surprising {\it unipolar} supersonic outflow.  The
flow has some resemblance to the convective envelope solution
discussed in this paper.  Only a small fraction of the circulating
matter is directly accreted by the black hole, and the energy required
to support the circulation is extracted from the infalling mass with
an efficiency $\sim 10^{-3}-10^{-2}$.  The circulation transports the
energy to large radii, where the author argues it could be radiated
via bremsstrahlung.  It would appear that in this model, viscosity is
large enough to suppress convection on scales $\lsim R$, but global
circulation on a scale $\sim R$ still survives, and this transports
energy.

The parallel between this solution and our convective envelope
solution is fairly strong.  The convective envelope again converts a
fraction of the binding energy of the accreting mass into an outward
flux of energy, which is carried by convection.  This energy somehow
has to be got rid of at the outer boundary.  One way in which the
energy could be eliminated from the system is by driving a slow (but
perhaps massive) outflow from the outer boundary.  Another is that the
energy could be radiated by gas on the outside, as suggested by
Igumenshchev (2000).

The latter is quite plausible.  Since $\rho\propto R^{-1/2}$ and
$T\propto R^{-1}$, the bremsstrahlung cooling rate per unit volume
varies as $Q_{brems}^-\propto\rho^2T^{1/2} \propto R^{-3/2}$.  The
cooling time scale thus goes as $t_{cool}\propto\rho
T/Q_{brems}^-\propto R^0$.  In comparison, since $v\propto R^{-3/2}$,
the accretion time scale goes as $t_{acc}=R/v\propto R^{5/2}$, and the
ratio $t_{acc}/t_{cool}\propto R^{5/2}$ increases very rapidly with
radius.  Therefore, one naturally expects a highly advection-dominated
convection zone (presumably well described by our convective envelope
solution) sandwiched between an inner energy producing zone close to
the black hole and an outer radiating zone.  The radiating zone would
have very specific spectral signatures which would be interesting to
explore.

Is the convective envelope solution relevant for real accretion flows?
This depends on the answers to two questions.

The first question is: What is the value of $\alpha$ in real ADAFs?
If $\alpha> \alpha_{crit2}\sim0.05$, then the convective envelope
solution is not possible and the second question below is irrelevant.
Empirical estimates of $\alpha$ from fitting the spectra of black hole
systems suggest that $\alpha$ is large.  Narayan (1996) obtained
$\alpha\sim1$, which Esin, McClintock \& Narayan (1997) later revised
to $\alpha\sim0.2-0.3$.  Quataert \& Narayan (1999) considered models
with $\alpha=0.1$, but they did not verify that such low values of
$\alpha$ are consistent with the constraints that Narayan (1996) and
Esin et al. (1997) considered, specifically the existence of luminous
black hole X-ray binaries in the ``low'' spectral state.  More work is
needed before one can state with confidence whether or not values of
$\alpha\lsim\alpha_{crit2}$ are consistent with observations.

On the theoretical front, although there is good reason to believe
that ``viscosity'' in differentially-rotating accretion flows is
produced by magnetic stresses generated by the Balbus \& Hawley (1992,
1998) instability, the theory has not developed to the point where the
value of $\alpha$ can be estimated.  Numerical MHD simulations give
values over a wide range, $\alpha\lsim0.01$ to $\alpha\sim0.6$ (Balbus
\& Hawley 1998).  Furthermore, all simulations so far have been done
on thin accretion disks, and it is unclear whether those results are
valid for the much thicker ADAFs.  Numerical MHD simulations of
accretion flows under ADAF-like conditions would be very worthwhile.
The simulations will need to be done in three dimensions rather than
two, since even in the thin disk case, there are large differences
between 2D and 3D simulations (Balbus \& Hawley 1998).

The second question is: Does convection in differentially-rotating
ADAFs move angular momentum outward or inward?

Ryu \& Goodman (1992) showed that linear modes in a convectively
unstable thin accretion disk transfer angular momentum inward.  Stone
\& Balbus (1996) used numerical simulations to study the non-linear
version of the problem.  They found that angular momentum was either
transported very weakly inward or not at all.  These are significant
results, but their relevance to ADAFs is a little uncertain.  As NY
emphasized, the entropy gradient in a thin disk model is in the
vertical direction whereas the angular momentum gradient is in the
horizontal direction.  In ADAFs on the other hand, both the entropy
gradient and the angular velocity/angular momentum gradient are in the
radial direction.  This might conceivably cause some differences in
the physics.

Kumar, Narayan \& Loeb (1995) analysed convective angular momentum
transport in a differentially rotating medium using a mixing length
formalism.  They found that, while small amplitude linear
perturbations transport angular momentum down the specific angular
momentum gradient, in agreement with the result of Ryu \& Goodman
(1992), nonlinear saturated convection generally behaves very
differently.  Indeed, in the nonlinear regime, the angular momentum
transport depends critically on the nature of the interactions between
convective eddies, which Kumar et al. modeled via a scattering term.
By tuning their scattering function, they could reproduce a wide
variety of behaviors, covering the entire range from $g=-1/3$ to $g>1$
(in the language of the present paper).

Stone \& Balbus (1996) analyzed the basic hydrodynamic equations in a
turbulent Keplerian disk and showed that, when azimuthal pressure
perturbations are small, there can be no net outward transport of
angular momentum.  As already mentioned, this powerful result is the
reason why axisymmetric numerical simulations of convecting disks
always move angular momentum inward.  It also throws considerable
light on the Kumar et al. (1995) work since the eddy scattering
function invoked by these authors implicitly involves azimuthal
pressure fluctuations.  It is the scattering that enables their model
to transport angular momentum outward.

Despite these theoretical studies, the basic question of how angular
momentum transport actually operates in a convection-dominated
accretion flow remains very much open.  Specifically, it is unclear
how important azimuthal pressure perturbations are in these systems,
and whether angular momentum is transported outward or inward.  We see
only one way to answer this question: full three-dimensional numerical
simulations.  If 3D simulations are too expensive, one should at least
carry out nonaxisymmetric two-dimensional simulations in cylindrical
geometry.

One empirical fact is worth noting; the convection zone in the Sun is
closer to being in a state of constant angular velocity than a state
of constant angular momentum.  Indeed, in the equatorial plane, the
angular velocity actually increases with increasing radius, which
motivated Kumar et al. (1995) to consider models with $g>1$.  One of
the two conditions, (i) and (ii), mentioned at the beginning of this
section must be violated in the Sun.  One possibility is that the
effective viscous $\alpha$ in the Sun is large enough to counter any
inward transport of angular momentum by convection.  This is
reasonable since the convective motions in the Sun are not very
energetic, and $\alpha$ does not have to be very large to counter the
effect of convection.  Alternatively, perhaps convection in the Sun
behaves as in equation (6) and drives the system to near-zero
$d\Omega/dR$.  The Sun is certainly not a good comparison for an ADAF
--- a more rapidly rotating convective star would be closer.
Nevertheless, the fact that the Sun has achieved an equilibrium
configuration that is very different from a constant specific angular
momentum state must be treated as a possible clue as we grope towards
a better understanding of advection-dominated flows.

In closing, we note that the key point of this paper is that there is
considerable uncertainty associated with angular momentum transport by
convection, and that this leads to two very distinct configurations
for an ADAF.  There is a related, though less uncertain, issue
connected to energy transport.  We considered in this paper convective
transport, where energy is transported down the entropy gradient.  In
addition, there could also be conduction which transports energy down
the temperature gradient.  Conduction has been studied by Gruzinov
(1999) in the context of Bondi accretion.  It would be of interest to
extend Gruzinov's analysis to ADAFs and to investigate models in which
convection and conduction are both present in an accretion flow.

\noindent{\it Acknowledgments.}  The authors thank Eliot Quataert and
Andrei Gruzinov for sharing their results prior to publication and for
permitting use of some of their methods, the referee, Mitch Begelman,
for useful comments, and SISSA (Trieste, Italy) for hospitality while
this work was done.  RN thanks Dimitar Sasselov for guidance on the
mixing length parameter $l_M$.  This work was supported in part by NSF
grants PHY 9507695 and AST 9820686, and the Royal Swedish Academy of
Sciences.

\bigskip\bigskip
{
\footnotesize
\StartRef
\noindent {\large \bf References} \\

\Ref Abbett, W., Beaver, M., Davids, B., Georgobiani, D., Rathbun, P.,
\& Stein, R. 1997, ApJ, 480, 395 \\
\Ref Abramowicz, M., Chen, X., Kato, S., Lasota, J.-P., \& Regev, O.,
1995, ApJ, 438, L37 \\ 
\Ref Abramowicz, M., Czerny, B., Lasota, J. P., \& Szuszkiewicz,
E. 1988, ApJ, 332, 646 \\
\Ref Asida, S. M. 1999, ApJ, in press (astro-ph/9907446) \\
\Ref Balbus, S. A., \& Hawley, J. F. 1992, ApJ, 400, 610 \\
\Ref Balbus, S. A., \& Hawley, J. F. 1998, RMP, 70, 1 \\
\Ref Begelman, M. C., \& Meier, D. L. 1982, MNRAS, 253, 873 \\
\Ref Blandford, R. D., \& Begelman, M. C., 1999, MNRAS, 303, L1 \\
\Ref Chen, X., Abramowicz, M. A., Lasota, J. P., Narayan, R., \&
Yi, I. 1995, ApJ, 443, L61 \\
\Ref Cox, J. P., \& Giuli, R. T. 1968, Principles of Stellar Structure
(New York: Gordon \& Breach) \\
\Ref Esin, A. A., McClintock, J. E., \& Narayan, R. 1997, ApJ, 489, 865 \\
\Ref Grossman, S. A., Narayan, R., \& Arnett, D. 1993, ApJ,
407, 284 (GNA)\\
\Ref Gruzinov, A. 1999, ApJ, submitted (astro-ph/9809265) \\
\Ref Honma, F. 1996, PASJ, 48, 77 \\
\Ref Ichimaru, S. 1977, ApJ, 214, 840 \\
\Ref Igumenshchev, I. V., Chen, X., \& Abramowicz, M. A. 1996, MNRAS, 
278, 236 \\
\Ref Igumenshchev, I. V., \& Abramowicz, M. A. 1999, MNRAS, 303, 309 \\
\Ref Igumenshchev, I. V. 2000, MNRAS, in press (astro-ph/9912170) \\
\Ref Kato, S., Fukue, J., \& Mineshige, S. 1998, Black-Hole Accretion
Disks (Kyoto: Kyoto Univ. Press) \\ 
\Ref Kato, S., \& Nakamura, K. E. 1998, PASJ, 50, 559 \\
\Ref Kim, Y.-C., Fox, P, Demarque, P., \& Sofia, S. 1996, ApJ, 461, 499 \\
\Ref Kippenhahn, R., \& Weigert, A. 1990 Stellar Structure and
Evolution (New York: Springer) \\
\Ref Kumar, P., Narayan, R., \& Loeb, A. 1995, ApJ, 453, 480 \\
\Ref Manmoto, T., Kato, S., Nakamura, K. E., \& Narayan, R. 2000,
ApJ, in press \\
\Ref Narayan, R. 1996, ApJ, 462, 136 \\
\Ref Narayan, R., Mahadevan, R., \& Quataert, E. 1998, in Theory of
Black Hole Accretion disks, eds. M. A. Abramowicz, G. Bjornsson, \&
J. E. Pringle (Cambridge Univ. Press), p148 \\ 
\Ref Narayan, R., \& Yi, I. 1994, ApJ, 428, L13 (NY) \\ 
\Ref Narayan, R., \& Yi, I. 1995a, ApJ, 444, 231 \\ 
\Ref Narayan, R., \& Yi, I. 1995b, ApJ, 452, 710 \\ 
\Ref Quataert, E., \& Gruzinov, A. 2000, ApJ, submitted 
(QG, astro-ph/9912440) \\
\Ref Quataert, E., \& Narayan, R. 1999, ApJ, 520, 298 \\
\Ref Ryu, D., \& Goodman, J. 1992, ApJ, 388, 438 \\
\Ref Stone, J. M., \& Balbus, S. A. 1996, ApJ, 464, 364 \\
\Ref Stone, J. M., Pringle, J. E., \& Begelman, M. C. 1999,
MNRAS, 310, 1002 \\
\Ref Tassoul, J.-L. 1978, Theory of Rotating Stars, Princeton
Univ. Press \\

\vfill\eject

\centerline{\bf Appendix A}

The model of convection which we employed in the main text of the
paper is somewhat limited because of our use of height-integrated
equations.  In particular, when we calculated the diffusion constant
associated with convection, we set the characteristic growthrate of
convective motions equal to $(-N_{eff}^2)^{1/2}$, where
$N_{eff}^2=N^2+\kappa^2$, and $N$ is the Brunt-V\"ais\"al\"a frequency and
$\kappa=\Omega$ is the epicyclic frequency (see eqs 3 and 4).  This
is appropriate for a one-dimensional flow in which convective blobs
experience displacements only in the equatorial plane.  However, as
QG have pointed out and as is well-known for rotating fluids (e.g.
Tassoul 1978, Begelman \& Meier 1982), in the full 2D flow the most
unstable modes are not in the equatorial plane.  In this Appendix we
present a modified analysis in which we identify the most unstable
mode and use its properties to estimate the effects of convection.
We draw heavily on the work of QG (specifically, the Appendix of
their paper).

We consider an axisymmetric self-similar 2D flow in polar
coordinates: $r\theta$.  We assume that the density varies as
$r^{-a}$ (QG use $n$ instead of $a$) and that the azimuthal velocity
$v_\phi$ and the sound speed $c_s$ vary as $r^{-1/2}$.  Using QG's
results as a guide, we assume that $v_\phi$ and $c_s$ at a given $r$
are both independent of $\theta$.  (QG showed this only for the
convectively marginally stable system, but we assume that the result
is valid also for the more general situation we consider.)  In terms
of the Keplerian velocity $v_K=(GM/r)^{1/2}$, we may write
$$
v_\phi(r,\theta)=\Omega_0 v_K(r). \eqno (A1)
$$

Consider a blob of material at some arbitrary $r\theta$ and imagine
imposing a unit displacement: $dr=\cos\chi$, $rd\theta=\sin\chi$.
Following the analysis presented by QG, the effective frequency
$N_{eff}$ associated with the dynamics of the blob is given by
$$
-{\gamma N_{eff}^2\over\Omega_K^2}=A\cos^2\chi+B\sin^2\chi
+C\cos\chi\sin\chi, \eqno (A2)
$$
where the quantities $A$, $B$ and $C$ are (we have translated the
expressions given in QG to the notation of this paper):
$$
A=1-a(\gamma-1)+[a(\gamma-1)-(\gamma+1)]\Omega_0^2, \eqno (A3)
$$
$$
B=-{(a+1)(\gamma-1)\cot^2\theta\Omega_0^4\over(1-\Omega_0^2)}
-2\gamma\cot^2\theta\Omega_0^2, \eqno (A4)
$$
$$
C=2[a(\gamma-1)-(\gamma+1)]\cot\theta\Omega_0^2. \eqno (A5)
$$
QG show that there are additional terms in $B$ and $C$ which depend on
$dv_\phi/d\theta$, but these terms vanish under the assumption that
$v_\phi$ is independent of $\theta$.

In the height-integrated problem considered in the main text, there is
a unique frequency associated with convective motions, namely
$N_{eff}=(N^2+\kappa^2)^{1/2}$.  In the more general problem
considered here, there is a multitude of frequencies; indeed, the
frequency is a function of two angles, the polar angle $\theta$
corresponding to the location of a blob, and the angle $\chi$
associated with the direction of displacement of the blob.  (The
analysis in the main paper corresponds to the specific choice
$\theta=\pi/2$, $\chi=0$.)  

In the two-dimensional space of $\theta$ and $\chi$, the value of
$-N_{eff}^2$ varies as a function of the two angles.  If $-N_{eff}^2
\leq0$ (i.e. $N_{eff}^2\geq0$) for all choices of the angles, then
the medium is convectively stable.  In this case, the convective
diffusion constant $K_c$ is zero and so is the dimenionless constant
$\alpha_c$.  We are not interested in such flows.  Rather, we are
interested in those flows in which $-N_{eff}^2>0$ for at least some
choices of $\theta$ and $\chi$.  If we determine the largest positive
value of $-N_{eff}^2$ in such a flow, then $(-N_{eff}^2)^{1/2}$ would
give the growthrate of the most unstable convective mode.  We assume
that this most unstable mode dominates the convection.

To determine the largest positive value of $-N_{eff}^2$, we first
differentiate (A2) with respect to $\chi$ and set the result to zero.
We find that the most unstable mode at a given $\theta$ corresponds
to displacements that satisfy
$$
\cos2\chi={A-B\over[(A-B)^2+C^2]^{1/2}},\qquad
\sin2\chi={C\over[(A-B)^2+C^2]^{1/2}}. \eqno (A6)
$$
Substituting this value of $\chi$ back in (A2) we obtain the
maximum value of $(-N_{eff}^2)$ at each $\theta$:
$$
\left(-{\gamma N_{eff}^2\over\Omega_K^2}\right)_{\theta, max}=
{1\over2}\left((A+B)+[(A-B)^2+C^2]^{1/2}\right). \eqno (A7)
$$
Maximizing this quantity next with respect to $\theta$, we obtain the
overall maximum value of $(-N_{eff}^2)$.  We find that for all flows
that are convectively unstable, i.e. which have $-N_{eff}^2>0$ over
at least some range of $\theta$ and $\chi$, the maximum positive value of
$-N_{eff}^2$ occurs at $\theta=0$.  Now, as $\theta\to 0$, $A$
remains finite, but $B$ and $C$ both diverge, with $B$ diverging
faster than $C$ ($\cot^2\theta$ versus $\cot\theta$).  This
simplifies matters, and so we find that
$$
\left(-{\gamma N_{eff}^2\over\Omega_K^2}\right)_{max}=
\left(A-{C^2\over4B}\right)_{\theta\to0}.
\eqno (A8)
$$
Both in (A7) and (A8), the corresponding mode is convectively
unstable only if the quantity on the right hand side is positive.
We focus only on such cases.

As noted by QG, the simple ``equatorial'' analysis presented in the
main text of this paper corresponds to using just $A$ in equation (A8).
The more complete analysis presented here shows that the
correct expression corresponds to the quantity $(A-C^2/4B)$.  This is
larger than $A$ because $B$ is negative.  Thus, for instance, it is possible
for $A$ to be negative --- which means that the medium is stable to
equatorial displacements --- and yet displacements at non-equatorial
values of $\theta$ could be unstable, so that the medium as a whole
could be convectively unstable.  To evaluate (A8), we note that
although $B$ and $C$ diverge as $\theta\to0$, the ratio $C^2/B$ does
not, and so $(-N_{eff}^2)_{max}$ remains finite as $\theta\to0$.  The
full expression is
$$
\left(-{\gamma N_{eff}^2\over\Omega_K^2}\right)_{max}=
1-a(\gamma-1)+[a(\gamma-1)-(\gamma+1)]\Omega_0^2
$$
$$
\qquad\qquad\qquad +{[a(\gamma-1)-(\gamma+1)]^2\Omega_0^2(1-\Omega_0^2)\over
2\gamma(1-\Omega_0^2)+(a+1)(\gamma-1)\Omega_0^2}. \eqno (A9)
$$

Having determined $(-N_{eff}^2)_{max}$, the rest of the analysis
proceeds exactly as in the main paper, except that we replace equation
(4) for the convective diffusion constant by
$$
K_c={L_M^2\over4}\left(-N_{eff}^2\right)_{max}^{1/2}. \eqno (A10)
$$
Note that this result is valid only if $(-N_{eff}^2)_{max}>0$;
otherwise, $K_c=0$.  Just as equation (4) is replaced by (A10),
equation (5) is also correspondingly modified.  None of the other equations
needs to be modified.  The radial momentum equation, for instance, is
still given by equation (1).  For NY-like self-similar flows with
$a=3/2$ we make use of equation (10), while for the convective
envelope solution with $a=1/2$ we use equation (12).

The dashed lines in Figs. 1 and 2 show some numerical results.  We
see that the results are qualitatively similar to what we obtained
with the simpler theory presented in the main text (shown by solid
lines in the plots).  The value of $\alpha_c$ is larger now compared
to the height-integrated theory, as one might expect.  The critical
$\alpha_{crit2}$ up to which the convective envelope solution is
possible is also larger by a few tens of per cent.  Interestingly,
the other critical $\alpha_{crit1}$ is unaffected.  When
$\alpha=\alpha_{crit1}$, the solution has no rotation and the flow is
perfectly spherical.  In this limit, equatorial displacements capture
the full story and the more detailed analysis presented here adds
nothing.

The problem considered by QG corresponds
to a convectively marginally stable accretion flow in which
$\alpha\to0$, $\alpha_c\to0$.  They show that in this limit the
Bernoulli parameter, 
$$
Be={1\over2}v_\phi^2+{1\over2}v^2+{\gamma\over\gamma-1}c_s^2-v_K^2,
\eqno (A11) 
$$ 
is zero.  We confirm this result.  The reader is referred to NY,
Narayan \& Yi (1995a) and Blandford \& Begelman (1999) for a
discussion of the significance of the Bernoulli parameter.  For
non-vanishing $\alpha$ and $\alpha_c$, we find that the Bernoulli
parameter is always positive.  For $\alpha=\alpha_{crit2}$, for
instance, we obtain $Be\sim0.3v_K^2$.

The positivity of the Bernoulli parameter for finite $\alpha_c$ in
the convective envelope solution is fairly easy to understand once we
realize certain properties of these flows.  First, in these flows
$v=0$, and $v_\phi$ and $c_s$ are independent of $\theta$ (according
to the analysis presented by QG which we have borrowed).  Therefore,
even though the flow rotates and is non-spherical, $Be$ is
nevertheless independent of $\theta$ and is a function only of $r$.
Second, the flows we are interested in are convectively unstable.  In
order to have a convective instability, the Bernoulli parameter has
to be a decreasing function of increasing radius.  Finally, the flows
are self-similar, which implies that $Be$ must be proportional to
$v_K^2(r)$.  The latter two conditions can both be satisfied only if
$Be$ is positive.  Therefore, for finite $\alpha_c$, we expect a
positive value of $Be$.  In the limit $\alpha_c\to0$ considered by
QG, the flow is marginally stable to convection.  This means that
$Be$ must be independent of $r$, which can be reconciled with
self-similarity only if $Be=0$ (as QG found).

\newpage
\vskip 5in
\newpage

\begin{figure}
\plotone{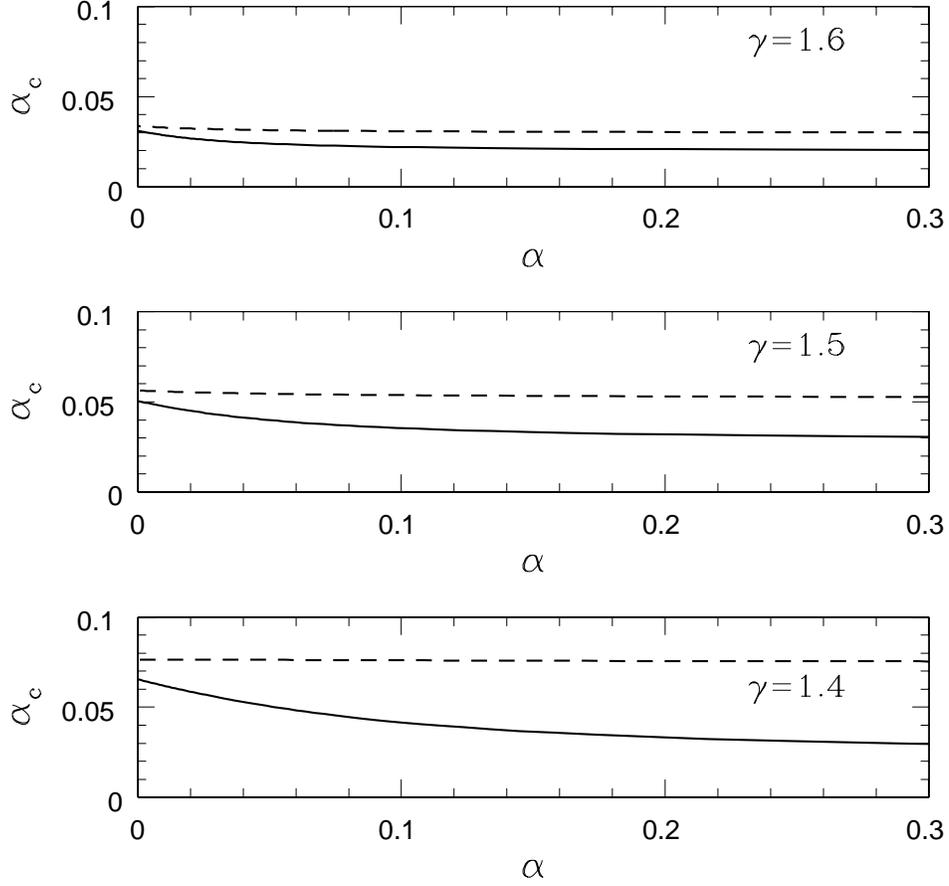}
\caption{Variation of the convective coefficient $\alpha_c$ as a
function of the viscosity coefficient $\alpha$ for three values of the
adiabatic index $\gamma$.  The calculations correspond to the case
when convection moves angular momentum outward according to equation
(6).  The solutions are of the self-similar form derived by NY, in
which the density scales as $\rho\propto R^{-3/2}$.  Note that there
is a consistent solution for all values of $\alpha$.  The solid lines
correspond to the height-integrated version of the theory described in
the main paper, and the dashed lines correspond to the analysis
described in the Appendix.}
\end{figure}

\newpage
\vskip 5in
\newpage

\begin{figure}
\plotone{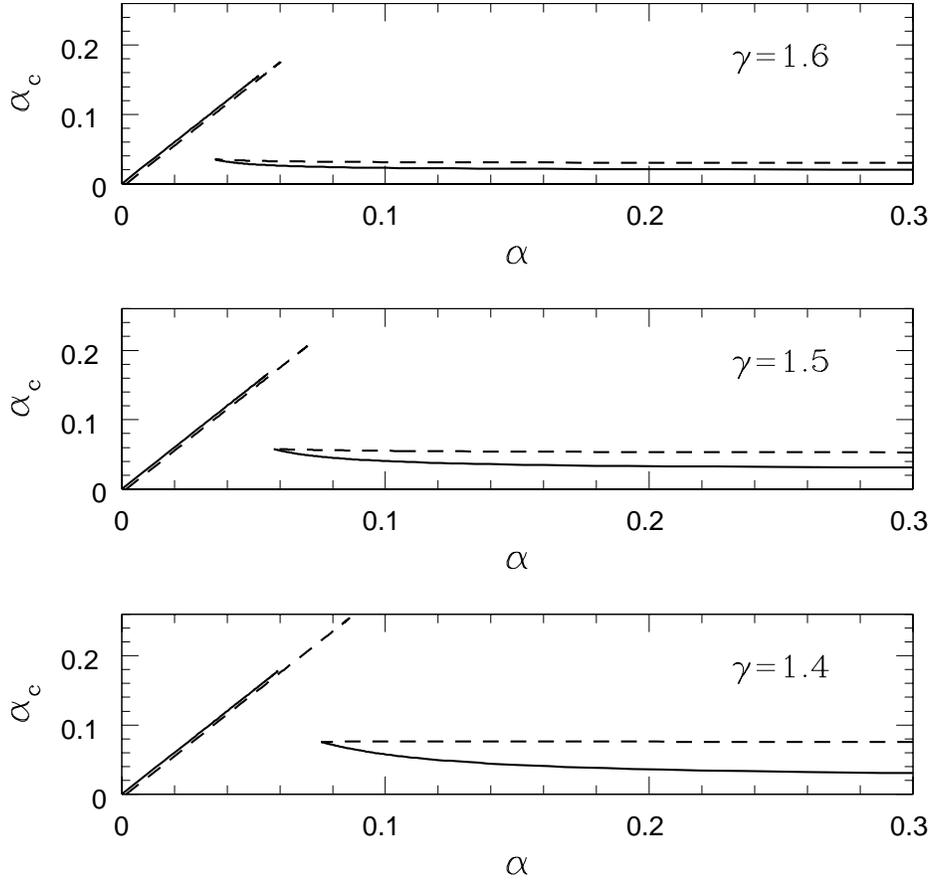}
\caption{Variation of the convective coefficient $\alpha_c$ as a
function of the viscosity coefficient $\alpha$ for three values of the
adiabatic index $\gamma$.  The calculations correspond to the case
when convection moves angular momentum inward according to equation
(7).  The lines on the right refer to a solution of the self-similar
form derived by NY, in which the density scales as $\rho\propto
R^{-3/2}$.  This solution is only available for $\alpha$ greater than
a critical value $\alpha_{crit1}$, which depends on $\gamma$.  The
lines on the left correspond to the convective envelope solution
discussed in \S5.2 in which $\rho\propto R^{-1/2}$.  This solution is
only available for $\alpha$ less than a critical value
$\alpha_{crit2}$, which again depends on $\gamma$.  The solid lines
correspond to the height-integrated version of the theory described in
the main paper, and the dashed lines correspond to the analysis
described in the Appendix.}
\end{figure}

\newpage
\vskip 5in
\newpage

\begin{figure}
\plotone{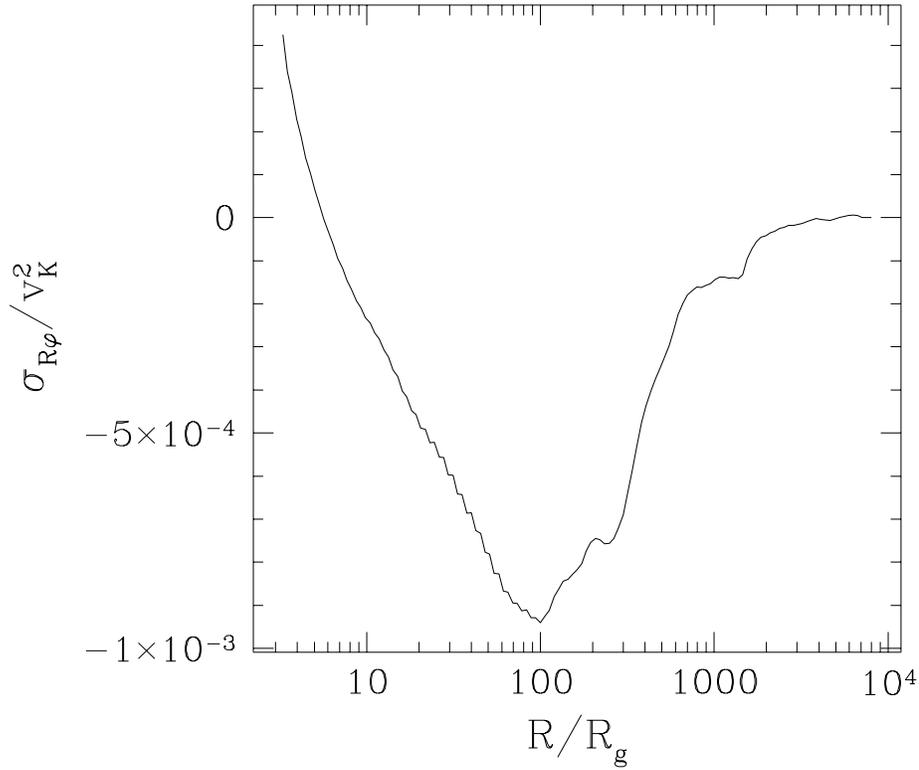}
\caption{Shows the $(R,\phi)$-component of the Reynolds stress tensor,
$\sigma_{R\phi}$, as a function of radius in an axisymmetric numerical
simulation of an ADAF with $\alpha=0.01$ and $\gamma=5/3$.  The
Reynolds stress has been averaged over the polar angle $\theta$ (as
explained in the text), and normalized to $v_K^2$.  Apart from two
regions near the boundaries, $R<5R_g$ and $R>5\times 10^3 R_g$, the
stress is negative.  This confirms that convection in axisymmetric
simulations moves angular momentum inward, as argued by Stone \&
Balbus (1996).}
\end{figure}

\newpage
\vskip 5in
\newpage

\begin{figure}
\plotone{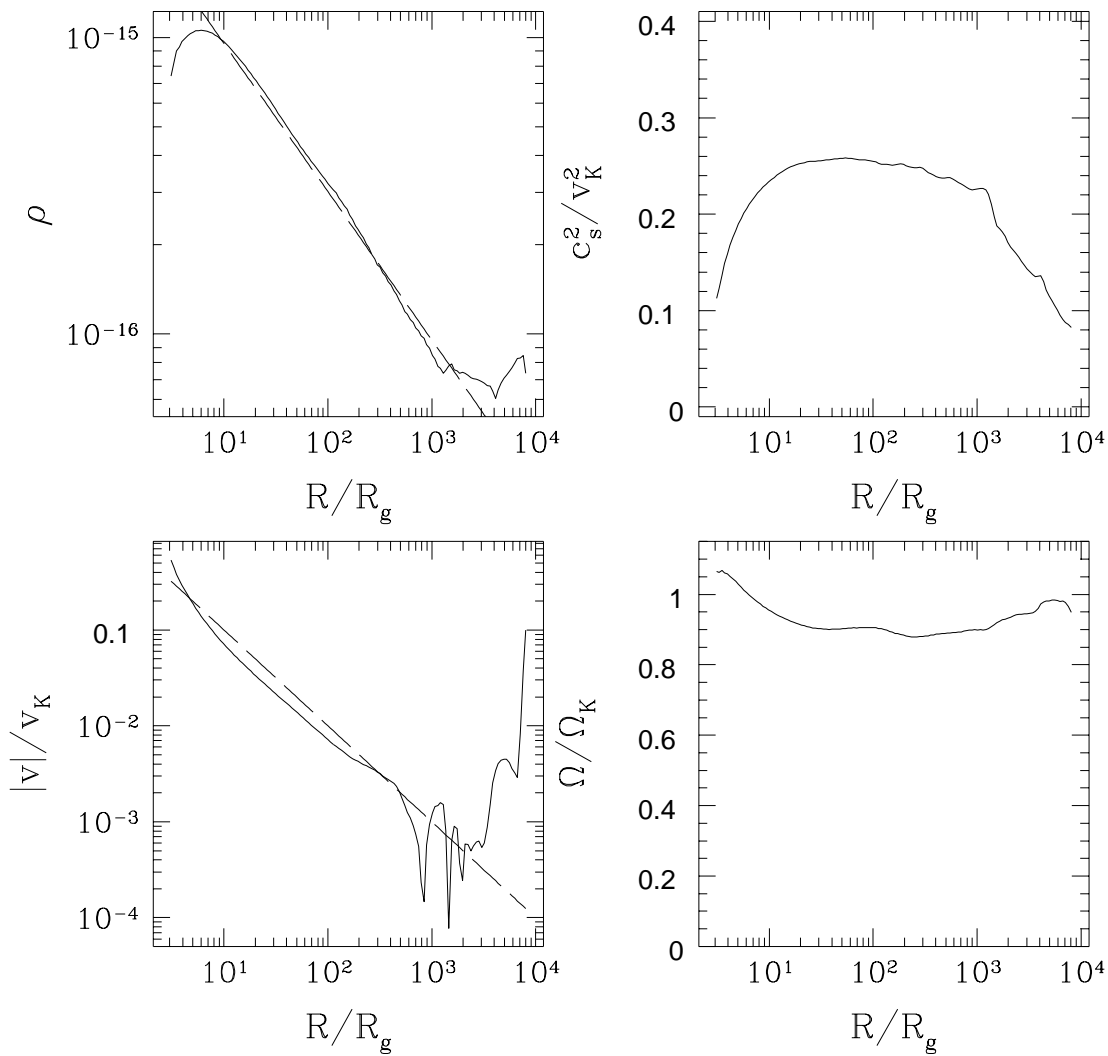}
\caption{Profiles of $\rho$ (solid line, upper left panel),
$c_s^2/v_K^2$ (upper right), $|v|/v_K$ (solid line, lower left) and
$\Omega/\Omega_K$ (lower right), shown as functions of radius, for an
axisymmetric convectively unstable accretion simulation with
$\alpha=0.01$ and $\gamma=5/3$.  All quantities have been averaged
over polar angle $\theta$ and time.  Except near the boundaries
($R<10R_g$ and $R>10^3 R_g$), the profiles of $\rho$, $c_s^2/v_K^2$
and $\Omega/\Omega_K$ show the power-law behaviors predicted by the
self-similar convective envelope solution; also the radial velocity
$v$ is small as predicted.  In the upper left panel the dashed line
corresponds to the analytical scaling $\rho\propto R^{-1/2}$.
Similarly, in the lower left panel the dashed line corresponds to the
predicted scaling $|v|/v_K\propto R^{-1}$.}
\end{figure}

\end{document}